*Original Article*

# Performance Evaluation of Parallel Algorithms

Donald Ene[1], Vincent Ike Anireh[2]

[1,2]*Department of Computer Science, Rivers State University, Port Harcourt, Nigeria*

**Abstract -** *Evaluating how well a whole system or set of subsystems performs is one of the primary objectives of performance testing. We can tell via performance assessment if the architecture implementation meets the design objectives. Performance evaluations of several parallel algorithms are compared in this study. Both theoretical and experimental methods are used in performance assessment as a subdiscipline in computer science. The parallel method outperforms its sequential counterpart in terms of throughput. The parallel algorithm's performance (speedup) is examined, as shown in the result.*

## 1. Introduction

Performance assessment aims to figure out how a computer works as a whole and as a collection of subsystems. We can determine whether an architecture's implementation meets the design goals by evaluating its performance [1]. Performance evaluation is a subdiscipline of computer science that includes theoretical models of machine behavior and experimental study of real hardware [2].

A concurrent system has a tremendous amount of complexity. Several processors are working at the same time, executing programs. Program execution success depends on communication and synchronization between individuals or groups of processors. Resource contention and information flow need to increase the system's total load. In addition, the parallel architecture's job division and balancing add to the overall system's load. A user must indicate how well a system performs a specific task, despite its enormous complexity. There are several advantages to using parallel processing, such as identifying bottlenecks and fine-tuning programs and architectures to do a job as quickly as feasible. [3].

The term "parallel algorithm" refers to a method that may be conducted in pieces on many processing devices and then reassembled to produce the desired outcome. This study will evaluate a few algorithms.

### 1.1. Characteristics of Parallel Algorithms
The following are the important characteristics of parallel algorithms:
1. Deterministic and non-deterministic: Only deterministic algorithms can be implemented on a real-world computer.
2. The granularity of a calculation is defined by the size of the data and software modules that go into it. There are three gradations of algorithmic granularity to consider: coarse, medium, and fine.
3. Parallelism profile: The distribution of parallelism (DOP) in an algorithm exposes the potential for parallel processing, as shown by DOP's distribution. A parallel algorithm's effectiveness is often compromised.
4. Communication patterns and synchronization requirements: Communication patterns address memory access and interpose communications between processors or Pes. The patterns can be static or dynamic, depending on the algorithms. The static algorithm is more suitable for single instruction multiple data (SIMD) or pipeline machines, while dynamic algorithms are more suitable for multiple instruction multiple data (MIMD) machines. The synchronization frequency often affects the efficiency of an algorithm.
5. Uniformity of operations refers to the basic procedures that must be carried out. Since efficiency is increased, SIMD processing or pipelining is more desired if the processes are homogeneous throughout the data set. MIMD machines are better suited to random-structured algorithms.
6. Memory requirements and data structures: when solving a large-scale problem, the data set may require huge memory space. Data structures are chosen, and data movement patterns in the algorithms can affect the efficiency of a program. Both time and space complexities are key measures of the granularity of parallel algorithms.

## 2. Literature Review

In 2013, Chopra suggested that it is possible to tackle a given issue concurrently and cooperatively using a collection of K constants processes. In a sequential algorithm, if K is one, it is termed sequential. In a task system, some points are there, where processes communicate with each other to maintain the integrity of a parallel algorithm [4]. Interaction points are the name given to these points. These points define stages. A process may communicate with others before moving on to the next computation step after every stage. Because of the interdependencies between processes, certain interactions may be inhibited. Algorithms in which certain processes must wait for the completion of others are known as synchronized algorithms. An individual process's execution time depends on input data and system interruption, and all processes must wait for the slowest among them to complete before they may proceed. The worst-case calculation speed may lead to slower speedup and CPU use than intended. The synchronized algorithm's main flaw is this [5].

To overcome the problems encountered by synchronized parallel algorithms, Aswin Posited that asynchronized parallel algorithms exist for the problems in which processes are not generally required to wait for each other. Communication is achieved by reading dynamically updated global variables stored in shared memory since concurrent memory access is performed. Conflicts may occur, introducing some small delay in processes accessing common variables. An algorithm which requires execution on the multiprocessor system must be decomposed into a set of processes to exploit parallelism, which can be done by static or dynamic decomposition [4].

Before executing a static decomposition procedure, the collection of processes and their precedence relationships are known. If the number of processes is small and their flexibility is constrained, static decomposition gives the prospect of minimal process communication. A collection of process changes during execution in a dynamic decomposition technique can adapt to variations in the execution time of a process graph efficiently, but only at the cost of high process communication and other design overheads [5].

## 3. Asynchronized Parallel Algorithms

Several global variables are available to all processes in these algorithms. Global variables and shared data are used to communicate between processes. Some global variables are read when a process step is complete [6]. Once a subset of global variables has been modified, a process may either go on to the next stage or end depending on the values read from the variables and outcomes from the previous stage. One benefit of an asynchronous algorithm over a synchronous parallel method is the lack of an explicit dependence between the processes. Instead of waiting for input, this technique allows processes to proceed or stop depending on the information stored in global variables. Herein, the processes may be blocked from entering critical sections, which can be considered as one of its disadvantages; for example: $x_{i+1} = x_i - f'(x_i)f(x_i)$

Here, an asynchronous iterative algorithm consists of two processes, $P_1$ and $P_2$ and three global variables are defined. Process $P_1$ updates variables $V_1$ and $V_3$ while process $P_2$ updates $V_2$. Its disadvantages are also there. First, note that critical sections are needed in the algorithms. Secondly, the speedup of the algorithms is quite limited.

## 4. Synchronized Parallel Algorithms

An algorithm with synchronized parallel processes is one in which some stages of one process's program are not active until another process has finished a certain stage of its program, as opposed to a simple parallel algorithm. Various synchronized primitives may be used to create the synchronization. For example, to complete the matrix, Z = A.B + (C x D). (I + G) by maximum decomposition. A new parallel algorithm may be constructed. The matrix is decomposed into three processes: P1, P2, and P3.

Several global variables are available to all processes in these algorithms. Global variables and shared data are used to communicate between processes. Some global variables are read when a process step is complete. Once a subset of global variables has been modified, a process may either go on to the next stage or end depending on the values read from the variables and outcomes from the previous stage. One benefit of an asynchronous algorithm over a synchronous parallel method is the lack of an explicit dependence between the processes. Rather than waiting for inputs, processes in this algorithm either continue running or are terminated based on the data stored in global variables. Herein, the processes may be blocked from entering critical sections, which can be considered as one of its disadvantages; for example:

$$x_{i+1} = x_i - f'(x_i)f(x_i) \qquad (1)$$

## 5. Speedup and Efficiency

An algorithm's speedup over a sequential algorithm is determined by comparing its sequential computation times to its parallel computation times. Speedup factors of n are referred to as n-fold. If a sequential strategy takes 10 minutes to calculate, whereas a parallel one takes just 2 minutes, we say there is a 5-fold speedup.

The observed speedup is the result of a combination of all implementation factors. Higher-end CPUs, for example, often equate to faster performance. Still, the speedup may be diminished if the processors are also used by other applications, such as parallel algorithms. If the task is ridiculously parallel, utilizing n-fold processors does not always result in n-fold speedup. A parallel algorithm's speedup is defined as the ratio of the rate at which an algorithm is performed (power) when a task is conducted on N processors to the rate at which it is done by just one. To keep things simple, the method to be completed will now be treated as an arbitrary task in this research. The speedup may therefore be defined in terms of the time necessary to finish this fixed algorithm on 1 to N processors (increase in power as a function of N). in the equation below, $T(N)$ is the amount of time it takes to execute the task on $N$ processors. Therefore, the ratio $S(N)$ is the speedup.

$$S(N) = \frac{T(1)}{T(N)} \qquad (2)$$

In many circumstances, the time $T(1)$ has a serial portion, $T_s$ and a parallelizable portion $T_p$, as previously mentioned. When the parallel component is broken up, the serial time does not decrease. The parallel time is reduced by a factor of $N$. As a result, a speedup can be expected as follows

$$S(N) = \frac{T(1)}{T(N)} = \frac{T_s + T_p}{T_s + T_p/N} \qquad (3)$$

The underlying machine's parallelism and the algorithm's implementation will define the idea of computing units. Cores, processors, and containers might be used to construct these units. The remainder of this chapter will presume that computing units are processors unless otherwise specified. It is a difficult choice to make since a parallel method may allow a variety of parallelisms.

On a given task, the efficiency of a parallel algorithm might be written thus

$$E = \frac{S_p}{P} \qquad (4)$$

Generally, $E \in [0, 1]$. Since $S_p$ may surpass $p$, and $E$ may be greater than 1. Therefore, we must consider a specific case of problem P to compute the speedup or efficiency. Considering all instances are identical in size, we'll assume they'll be completed in exactly the same length. Consider the instance when P is the sum of two dense square matrices multiplied together. Assume two problem cases $A, B, C, and\ D$, where $A, B, C, and\ D \in \mathbb{R}^{n \times n}$. The size of the first instance is equal to the sum of A's elements plus B's elements, which is $2n^2$. The second instance has a size of $2n^2$. As a result, $A \times B$ and $C \times D$ both have the same amount of work. This conclusion is supported by the fact that both instances will need the same number of floating-point operations to process.

## 6. Analysis of Speedup And Efficiency
Analysis of parallel algorithms relies heavily on asymptotic analysis. It is important to understand how the parameter values of a measure change when the metric itself becomes infinite. Speedup is the topic of our first asymptotic analysis model. As the number of processors and issue size increase, it aims to show how speedup changes. The parallel algorithm may be studied theoretically or empirically using this paradigm. The following section will follow theoretical considerations based on Amdahl and Gustafson's laws.[7]

As seen in Figure 1, the larger the task, the better the speedup we may get with several processors. As *n* grows, the speedup curve approaches the identity line ($y = x$). We get a speedup around *p* and, as a result, an efficiency near 1. We conclude that this is a scenario where the parallel technique may be used. In general, an algorithm may be considered scalable if it maintains its efficiency as the size and number of processors increase. In this case, we perform a projection of the speedup to show that it remains close to the identity function for high alues of *n* and *p*.

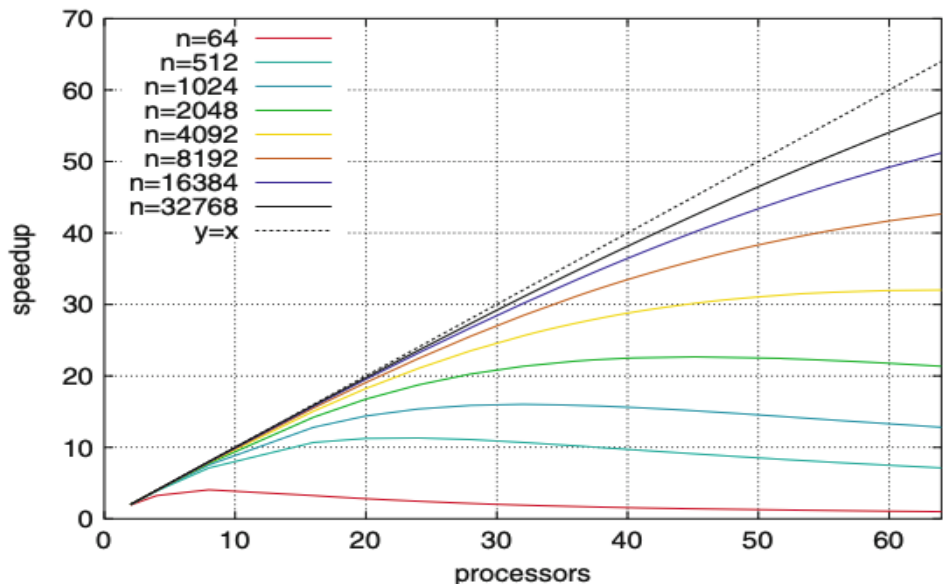

**Fig. 1 speedup distribution for the maximum element of a vector of different sizes of n**

## 7. Isoefficiency

For example, an "isoefficiency function" shows how increasing the number of processors in a parallel method does not affect the overall efficiency rating. No two parallel algorithms will share the same isoefficiency function for a given efficiency number. It is critical to distinguish between algorithms that expand exponentially as the number of processors rises (poorly scalable algorithms) and those that grow linearly as the number of processors increases (linearly scalable algorithms) (highly scalable algorithm). We estimated alternative efficiency values using the running time function from the "Asymptotic Analysis of Speedup and Efficiency" section, assuming that the issue size per processor (denoted by $n_p$) is 64 and 1024. If a linear speedup is readily obvious for np = 1024, it is not the case for np = 64, as shown. The topic of how to increase the problem size per number of processors is crucial in scalability analysis. For this reason, the idea of isoefficiency was developed [8].

The iso efficiency function has no known general method for calculation. However, such a function might be constructed if The execution time of a sequential and a parallel algorithm may be modelled analytically. Take, for example, a basic sequential algorithm with a running time of $T1 = n\, tc$, where n is the issue size, and $tc$ is the processing time. It is a simple scenario in which the sequential algorithm consists of performing a given operation n times.

## 8. Scalability Laws

In the last section, we showed how to use speedup and efficiency to test the scalability of a parallel method. In this part, how to calculate these measures hypothetically will be the focus.

### 8.1. Amdahl's Law

According to Amdahl's Law, many tasks are instantly ruled out of consideration for parallelization. Suppose the serial fraction of the code is not much less than the portion that might be parallelized. In that case, we will not notice significant speedup no matter how many nodes or how fast our connections are (if we rewrote it and were fortunate to split it up among nodes to complete in less time than it would otherwise). Despite this, Amdahl's Law remains too optimistic. It does not account for the overhead imposed because of parallelizing the function [9]. Amdahl's Law is a formula for calculating the maximal speedup from a part-sequential, part-parallel program. The Law calculates the total speedup, considering that the sequential composition of the program has no speedup.

In contrast, the parallel portion has a speedup S. It may appear unexpected that we only get a 3-fold overall speedup while 90% of the algorithm gets a 4-fold speedup. According to Amdahl's Law, the non-parallelizable section of the algorithm has a disproportionate influence on the total speedup. The formula of the Law states thus;

$$overall\ speedup = \frac{1}{(1-P)+\frac{P}{S}} \quad (5)$$

The percentage of the algorithm that can be parallelized is referred to as P. That method component has a speedup factor of S owing to parallelization. The following computation proves that Amdahl's Law is correct. Let, $T_s$ denote the computation time without parallelism and, $T_p$ denote the computation time with parallelism. The parallelism-induced speedup is then calculated as

$$total\ speedup = \frac{T_s}{T_p} \quad (6)$$

The value P in Amdahl's Law is the proportion of $T_s$ that can be parallelized, a number between 0 and 1. Then, the proportion of $T_s$ that cannot be parallelized is 1-P. Therefore,

$$T_p = time\ spent\ in\ unparallelizable\ code + time\ spent\ in\ parallelizable\ code = (1-P) \times T_s + P \times \frac{T_s}{S} \quad (7)$$

Now it can be concluded that $total\ speedup = \frac{T_s}{T_p} = \frac{T_s}{(1-P) \times T_s + P \times \frac{T_s}{S}} = \frac{1}{(1-P)+\frac{P}{S}}$ (8)

Consider the problematic assumptions that underline Amdahl's rule. The speedup achieved by this approach is at most linear. However, super-linear speedups have been seen. Amdahl's rule suggests a parallel algorithm is formed when sequential instructions are parallelized. On the other hand, the parallel technique may be derived from a completely different issue design [10].

### 8.2. Limits of Scalability

The P-completeness hypothesis, as presented by Stephen Cook, has one goal: to find fundamentally sequential problems. It indicates that there isn't a good parallel algorithm for resolving them. Greenlaw, Hoover, and Ruzzo present a compilation of P-complete problems [11] in their book, which contains several famous issues such as scheduling, minimal set cover, and linear programming. The memory wall is another key scaling constraint. A mismatch between memory bandwidth, latency, and processor speed causes the memory wall [12]. On certain computers, the time it takes to complete a DRAM Load/Store operation surpasses the time it takes to perform a multiplication. Numerous approaches have been used in computer machines to prevent such a stumbling block. The third constraint is the amount of energy consumed. The power consumption of supercomputers increases as the number of processors increases. This absorbed energy is converted to heat, which must be released. According to some studies, cooling may account for up to 40% of the energy utilized in a data center [13].

*8.3. Gustafson's Law*

Because of the idea of serializable and parallelizable work, Gustafson's Law is similar to Amdahl's (the global work is denoted by W as before). On the other hand, Gustafson's method presupposes that these proportions are known simultaneously rather than assessed sequentially. Assume that the serial and parallel fractions for a parallel algorithm running on p processors are $f'_{seq}$ and $f'_{par}$ respectively. The running time of the algorithm is as follows;

$$T_p = \left(f'_{seq} + f'_{par}\right) \cdot W \cdot t_c \qquad (9)$$

The running time for the corresponding sequential algorithm is

$$T_1 = \left(f'_{seq} + f'_{par} \cdot P\right) \cdot W \cdot t_c \qquad (10)$$

This leads up to the scaled speedup equation below

$$S_p = P + (1 - P) f'_{seq} \qquad (11)$$

Gustafson's work [14] is largely responsible for the modern understanding of scalability. Indeed, the original Amdahl study demonstrated that parallelization will always hit a limit given a constant issue size. This viewpoint is known as the strong scaling perspective today. Gustafson demonstrated that large parallel machines are not worthless without contradicting Amdahl's findings. Indeed, the more resources available, the faster huge issues can be solved. As a result, he presented the weak scaling analysis and the concept of measuring the algorithm's efficiency as issue sizes, and processor counts increased. Gustafson's study also reviewed Amdahl's analysis to illustrate how huge speedups in parallel algorithms may be achieved [15].

## 9. Conclusion

Multiprocessor efficiency relies on the performance aspect of an algorithm that is synchronized and iterative. Tightly connected systems suffer from memory interference. The cost of interprocess communication is the most important consideration in loosely linked systems. Synchronization and interprocess communication affect the performance of a synchronized iterative algorithm.

Sequential and parallel algorithms to solve linear equations of varying lengths are in Table 1, and the speedup is shown. The parallel method outperforms its sequential counterpart in terms of throughput. The parallel algorithm's performance (speedup) is examined by plotting the graph. As can be seen, the parallel approach is more efficient and saves considerable amounts of time during execution. In general, parallel algorithms are twice as fast as their sequential counterparts.